\documentclass[pra,showpacs,twocolumn,nofootinbib]{revtex4}
\usepackage{dcolumn}% Align table columns on decimal point
\usepackage{bm}% bold math
\usepackage{amssymb}
\usepackage{amsmath}
\usepackage{amsfonts}
\usepackage[]{graphicx}
\begin{document}
\bibliographystyle{prsty}
\title{ Can quantum mechanics be considered as statistical? an analysis of the PBR theorem  }
\author{Aur\'{e}lien Drezet}
\affiliation{Institut N\'eel UPR 2940, CNRS-University Joseph Fourier, 25 rue des Martyrs, 38000 Grenoble, France}

\date{\today}

\begin{abstract}
The answer to this question is `yes it can!' as we will see in
this manuscript. More, precisely after a discussion of
M.~F.~Pusey, J.~Barrett and T.~Rudolph (PBR) result
(arXiv:1111.3328) we will show that contrarily to the PBR claim
the epistemic approach is in general not disproved by their
`no-go' theorem.
\end{abstract}

\maketitle

\section{A $\Psi$-losophical introduction}
\subsection{Liouville's realm}
\indent Classical physics which is based on physical realism  makes the distinction
between ontic and epistemic state in a clean way. The `ontic'
state are the actual values of the dynamical variables $q(t),p(t)$
defined in the evolution or configuration space and solutions of the Hamilton or
Lagrange equations describing the system. They represent the
system even if there is not observer at least in a local universe (for a nonlocal universe where correlations can exist between disconnected region of space and time the definition should probably be amended a bit: one could for example states that the absence or presence of the observer should not disturb `too much' the rest of the universe).
The inclusion of the observer involves others dynamical variables
$Q(t),P(t)$ (therefore the observer is included in the theory). In principle, the coupling between the observer and
the system of interest could be reduced at will and therefore the
ontic state is also experimentally accessible. The `epistemic'
state is the density of probability $\rho(q,p,t)$ defined in the
same configuration space and which evolves in time following the
Liouville equation $d\rho(q(t),p(t),t)/dt=\partial\rho/\partial t+\{\rho,H\}=0$. It represents the
objective-subjective knowledge of the experimenter and is
statistic by nature (of course certainty is a particular
degenerate case of this general frame). In classical physics the
density being given at one time $t_0$  one can calculate it at any
times (past or future).  Additionally,  we can arbitrary
`mathematically' impose $\rho(t_0)$ in the equations.
Fundamentally, this means that the dynamic is decoupled from the
probabilistic evolution: the trajectories in the evolution space
are the same whatever the density function $\rho$ chosen.  This is
an important property which in part explains why the statistical
mechanics of Boltzmann-Gibbs requires some additional postulates
(based on symmetries  or plausible boundary conditions in the
remote past) in order to fix the equilibrium states of statistical
thermodynamics. The foundations of statistical physics is still a
subject of active research (in particular if we consider the
subjective-objective dualism concerning interpretation of
probability). However, its foundation relying on an ontic state
$q(t), p(t)$ is universally accepted  by classical physicist and
therefore never contradicts realism.
\subsection{Heisenberg's realm}
\indent In  quantum mechanics the situation is different. Indeed, we start
from a statistical theory `the epistemic state'  but we don't have
any dynamic or trajectory $q(t),p(t)$. Instead, we have
observable $Q,P$ which can not all be measured `simultaneously' for
the same individual system. This leads to the principle of
complementarity which states that measurement associated with non
commuting operator require experimental procedures which mutually
exclude each other. In the same vain by a generalization of
Heisenberg uncertainty principle we deduce that due to
entanglement, i.e. quantum correlation, with the measurement
apparatus we cannot define unambiguously et univocally the
hypothetical `classical' path followed by a particle in an
interferometer. Therefore, the wave-particle dualism cannot be
solved experimentally and the concept of trajectories become
somehow metaphysical. The introduction of hidden dynamical
variables written generically $\lambda(t)$ after Bell is therefore
regarded by most orthodox quantum practitioners as a kind of
useless superstructure identical by nature to the hypothetical
Ether postulated in the XIX$^{th}$ century. However, postulating
the mere existence of such $\lambda(t)$ has at least the advantage to solve the
problem of the `Heisenberg-cut' that is the duality
classic-quantum or observer-object which is so important in Bohr
philosophy\footnote{There is an additional problem with orthodox quantum mechanics not so much discussed: it concerns the concept of probability. Indeed, for a classical or quantum realist a probability for an event $\alpha$ is a frequency  of occurrence defined as the limit $lim_{N\rightarrow+\infty} n_{\alpha}/N$.  This is of course a postulate in the same sense as we postulate Newton's laws (indeed we can never experience infinity: this is also a reply to Bayesianism: a natural law is an hypothesis therefore we don't need to use a non-frequentist approach to probability). However,  since it requires `$N=\infty$'  we admit that probability is only a approximate tool used for practical reasons (ignorance for example).  It cannot be fundamental and can not be used as a final truth. The same should be true in quantum mechanics and therefore the theory can not be complete (I took and deliberately deviate this reasoning from C. Fuchs `QBism' interpretation).}. In the Copenhagen interpretation we must indeed accept
a form a macro realism (with all what this implies) together with
a micro `non-realism' (whatever this can mean). However, since the
cut is movable it is difficult to understand  how realism can mute
into non-realism or reciprocally (the `cut' leads even to
uncountable difficulties  if we consider seriously Einstein's
relativity and its arbitresses concerning space-time foliations
and reference frames). Clearly, if we accept the hidden variable
approach the problem is automatically solved in a simple and
drastic way since then the paradoxical cut does not exist anymore
(I think that it was also the point stressed by Schrodinger in its
famous cat example). I am not sure that practitioners of orthodox
quantum mechanics would really appreciate this fact. For them the
counter intuitive nature of such $\lambda$-theories (in particular
after Bell theorem concerning nonlocality in the 1960's) would
make the price too high to pay and they would probably prefer to
let the question open or at least not decidable. I would even say
than  in order to convince quantum mechanics practitioners one or
more revolutionary principles are  clearly missing to solve the
problem of nonlocality in a not ad-hoc way. Additionally, such a
model should ultimately make new predictions going beyond current
quantum mechanics (again the problem of Ether). \\
\indent Still, for the
present days it is at least on a logical ground remarkable that
hidden variable models can be precisely defined.  It was indeed in
my opinion the clear merit of de Broglie and Bohm to construct
such a hidden variable model (the only one which is working fine
for all practical purpose i.e. without modifying Schrodinger
equation I would even say). The model is classical in the ontic
sense discussed before since it introduces trajectories but it is
also epistemic since it reproduces every statistical predictions
of standard quantum mechanics through a clearly (unfortunately)
nonlocal and contextual dynamic.  For this last reason
it would be better to call the model neo-classic since there is no nonlocal interaction in classical XIX$^{th}$ century physics.\\
After Bell's work people get more interested in this topic and in
Bohm's work since they discovered that they can put some
experimental limits on the apriori infinite number of possible
$\lambda$-theories by using some `simple' no-go theorems. In
particular, local causal models can be eliminated if we reject
loopholes, fatalistic and superdeterministic approaches. By the
same approach
non-contextuality was also eliminated by Bell, Kochen and Specker (BKS).
\subsection{New no-go games? }
\indent Recently, a new work by M.~Pusey, J.~Barret and T.~Rudolph (PBR in
the following)  was put on arxiv \cite{PBR} and submitted for publication
claiming a new revolutionary no-go theorem.  This of course
stirred much debates in blog discussions (see for example the blog of M.~Leifer: http://mattleifer.info/ from which I stole the title of the present text) and Nature even posted an
article about it.  The idea of the PBR theorem will be discussed
in details below but shortly its aim can be summarized in a simple
way. Indeed, PBR show that if a hidden variable exists it can not
be epistemic in a specifical sense of the word epistemic. More
precisely, the theorem (which is I think mathematically true)
states that the only way to include hidden variable in a
description of the quantum world is to suppose that for every pair
of  quantum states $\Psi_1$ and $\Psi_2$ the density of
probability must satisfy the condition of non intersecting support
in the $\lambda$-space:
\begin{eqnarray}
\rho(\lambda,\Psi_1)\rho(\lambda,\Psi_2)=0  & \forall \lambda.
\end{eqnarray}
If this theorem is true it would really make hidden variables
redundant (as I perceived  it) since it could be possible to
define a bijection or relation of equivalence between the lambda
space and the Hilbert space: (loosely speaking we could in
principle make the correspondence $\lambda\Leftrightarrow\psi$).
Therefore it would be as if $\lambda$ is nothing that a new name
for $\Psi$ it self (not even an Ether). \\
Very recently I read the PBR paper
with a lot of interest in particular because I had the feeling
that they missed something. I will try in the following to show
what they missed and what it means really for hidden variable
theories. At the end I hope that I will manage to convince you
that it is still possible to deny the validity of Eq.~1 for most
interesting $\lambda$-models.
\section{The PBR theorem}
\subsection{orthogonal states}
We consider a simple Q-bit space $\mathbb{E}$ and two states $|\Psi_1\rangle$ and $|\Psi_2\rangle$ such that in the orthogonal basis $|\pm\rangle$ we have
\begin{eqnarray}
\langle +|\Psi_1\rangle=\langle -|\Psi_2\rangle=0.
\end{eqnarray} Clearly the states are orthogonal since
\begin{eqnarray}
\langle \Psi_2|\Psi_1\rangle=\langle \Psi_2|[|+\rangle\langle+|+|-\rangle\langle-|]\Psi_1\rangle\nonumber \\=\langle \Psi_2|+\rangle\langle+|-\rangle\langle-|\Psi_1\rangle=0.
\end{eqnarray}
We now consider a hidden variable model and we write the
probabilities to find the outcomes $\pm$
\begin{eqnarray}
|\langle +|\Psi_1\rangle|^2=P(+|\Psi_1)=\int \xi(+|\lambda)\rho_1(\lambda)d\lambda=0\nonumber\\
|\langle -|\Psi_2\rangle|^2=P(-|\Psi_2)=\int
\xi(-|\lambda)\rho_2(\lambda)d\lambda=0.
\end{eqnarray}
In these equations we introduced the conditional `transition' probabilities $\xi(\alpha|\lambda)$
for the outcomes $\alpha=\pm 1$ supposing given the hidden state
$\lambda$. We have of course $\xi(+|\lambda)+\xi(-|\lambda)=1$.
For the case here considered we deduce $\xi(+|\lambda)=0$ if $\rho_1(\lambda)\neq0$ and similarly  $\xi(-|\lambda)=0$ if $\rho_2(\lambda)\neq0$.\\
We then obtain that if $\rho_2(\lambda)\cdot\rho_1(\lambda)\neq0$ for some values of $\lambda$  (which means that $\rho_1$ and $\rho_2$ have intersecting supports in the $\lambda$-space ) then  $\xi(+|\lambda)=\xi(-|\lambda)=0$ for such $\lambda$ values. Now  this is impossible since we have by definition $\xi(+|\lambda)+\xi(-|\lambda)=1$ for every $\lambda$. We conclude therefore that $\rho_2(\lambda)\cdot\rho_1(\lambda)=0$ for every $\lambda$ i.e. that $\rho_1$ and $\rho_2$ have nonintersecting supports in the $\lambda$-space. \subsection{non-orthogonal states}
We consider in the same Q-bit space  the two states $|\Psi_1\rangle$ and $|\Psi_2\rangle$ defined by
\begin{eqnarray}
|\Psi_1\rangle=|0\rangle, & \textrm{and }|\Psi_2\rangle=|+\rangle
\end{eqnarray} where $|0\rangle$ and $|1\rangle$ is an orthogonal basis and where $|\pm\rangle=\frac{1}{\sqrt{2}}[|0\rangle\pm|1\rangle]$ is a second orthogonal basis.\\
Now we introduce the 2 Q-bit Hilbert space  $\mathbb{E}\otimes\mathbb{E}$ and the orthogonal basis
\begin{eqnarray}
|\Phi_1\rangle=\frac{1}{\sqrt{2}}[|0\rangle\otimes|1\rangle+|1\rangle\otimes|0\rangle]\nonumber\\
|\Phi_2\rangle=\frac{1}{\sqrt{2}}[|0\rangle\otimes|-\rangle+|1\rangle\otimes|+\rangle]\nonumber\\
|\Phi_3\rangle=\frac{1}{\sqrt{2}}[|+\rangle\otimes|1\rangle+|-\rangle\otimes|0\rangle]\nonumber\\
|\Phi_4\rangle=\frac{1}{\sqrt{2}}[|+\rangle\otimes|-\rangle+|-\rangle\otimes|+\rangle]
\end{eqnarray}
We are interested in the four states $|\Psi_1\rangle\otimes|\Psi_1\rangle$, $|\Psi_1\rangle\otimes|\Psi_2\rangle$, $|\Psi_2\rangle\otimes|\Psi_1\rangle$, and $|\Psi_2\rangle\otimes|\Psi_2\rangle$. We get the following coefficient matrix in the $\Phi$ basis:
 \begin{table}[h]
\begin{tabular}{l||c|c|c|c}
  & $|\Phi_1\rangle$ & $|\Phi_2\rangle$ & $|\Phi_3\rangle$ & $|\Phi_4\rangle$\\
\hline
\hline
$|\Psi_1\rangle\otimes|\Psi_1\rangle$ & 0 & $1/2$ & $1/2$ & $1/\sqrt{2}$\\
\hline
$|\Psi_1\rangle\otimes|\Psi_2\rangle$ & $1/2$ & 0 & $1/\sqrt{2}$ & $1/2$ \\
\hline
$|\Psi_2\rangle\otimes|\Psi_1\rangle$ & $1/2$ & $1/\sqrt{2}$ & 0 & $1/2$ \\
\hline
$|\Psi_2\rangle\otimes|\Psi_2\rangle$ & $1/\sqrt{2}$ & $1/2$ & $1/2$ & 0 \\
\hline
\end{tabular}
\caption{Coefficient table in the $\Phi$ basis. }
\end{table}\\
We now introduce a hidden variable model and we write the
probabilities $P(\Phi_i|\Psi_j\otimes\Psi_k)=|\langle
\Phi_i|\Psi_j\otimes\Psi_k \rangle|^2$ as
\begin{eqnarray}
P(\Phi_i|\Psi_j\otimes\Psi_k)=\int\int \xi(\Phi_i|\lambda,\lambda')\rho_j(\lambda)\rho_k(\lambda')d\lambda d\lambda'
\end{eqnarray}
where $i=[1,2,3,4]$ and $j,k=[1,2]$. In this PBR model there is a independence criteria at the preparation since we write $\rho_{j,k}(\lambda,\lambda')=\rho_j(\lambda)\rho_k(\lambda')$. The measurement is however obviously non local from the form of $\Phi_i$.\\
Now, clearly from the table we get:
\begin{eqnarray}
P(\Phi_1|\Psi_1\otimes\Psi_1)=\int\int \xi(\Phi_1|\lambda,\lambda')\rho_1(\lambda)\rho_1(\lambda')d\lambda d\lambda'=0\nonumber\\
P(\Phi_2|\Psi_1\otimes\Psi_2)=\int\int \xi(\Phi_2|\lambda,\lambda')\rho_1(\lambda)\rho_2(\lambda')d\lambda d\lambda'=0\nonumber\\
P(\Phi_3|\Psi_2\otimes\Psi_1)=\int\int \xi(\Phi_3|\lambda,\lambda')\rho_2(\lambda)\rho_1(\lambda')d\lambda d\lambda'=0\nonumber\\
P(\Phi_4|\Psi_2\otimes\Psi_2)=\int\int \xi(\Phi_4|\lambda,\lambda')\rho_2(\lambda)\rho_2(\lambda')d\lambda d\lambda'=0.\nonumber\\
\end{eqnarray}
The first line implies $\xi(\Phi_1|\lambda,\lambda')=0$ if
$\rho_1(\lambda)\rho_1(\lambda')\neq 0$. This condition is always
satisfied if $\lambda$ and $\lambda'$ are in the support of
$\rho_1$  in the $\lambda$-space and $\lambda'$-space. Similarly
the fourth line implies  $\xi(\Phi_4|\lambda,\lambda')=0$ if
$\rho_2(\lambda)\rho_2(\lambda')\neq 0$ which is again always
satisfied if $\lambda$ and $\lambda'$ are in the support of
$\rho_2$  in the $\lambda$-space and $\lambda'$-space.\\ Finally
the second and third lines imply  $\xi(\Phi_2|\lambda,\lambda')=0$
respectively $\xi(\Phi_3|\lambda,\lambda')=0$  if
$\rho_1(\lambda)\rho_2(\lambda')\neq 0$ respectively
$\rho_1(\lambda)\rho_2(\lambda')\neq 0$. Taken separately these
four conditions are not  problematic. However in  order to be true
simultaneously and then to have
\begin{eqnarray}
\xi(\Phi_1|\lambda,\lambda')=\xi(\Phi_2|\lambda,\lambda')=\xi(\Phi_3|\lambda,\lambda')=\xi(\Phi_4|\lambda,\lambda')=0\nonumber\\
\label{e6}
\end{eqnarray}
for a same pair of $\lambda,\lambda'$ the conditions require that
the supports of $\rho_1$ and $\rho_2$ intersect. If this is the
case Eq.~9 will be true for any pair $\lambda,\lambda'$ in the
intersection.\\
However, this is impossible since we must have
$\sum_{i=1}^{i=4}\xi(\Phi_i|\lambda,\lambda')=1$ for every  pair
$\lambda,\lambda'$.   We conclude that
$\rho_1(\lambda)\rho_2(\lambda)= 0$ i.e. the supports of $\rho_1$
and $\rho_2$ are disjoints.\\
The result is not yet completely general  since we studied only
two particular states of $\mathbb{E}$. In order to generalize this
result PBR considered  the pair of non orthogonal states
$|0\rangle$, $|0\rangle+\tan(\theta)e^{i\chi}|1\rangle$ (with
$0<\theta<\pi/2$ and $\chi$ a phase). Using a basis rotation by an
angle $\theta/2$ and absorbing the phase $\chi$ in the basis
definition this pair of states can be re-parameterized as
$|\psi_0\rangle=\cos{(\theta/2)}|0\rangle-\sin{(\theta/2)}|1\rangle$,
$|\psi_1\rangle=\cos{(\theta/2)}|0\rangle+\sin{(\theta/2)}|1\rangle$.\\
Next PBR considered the n-uplet states $|\Psi(x_1,...,
x_{n})\rangle$ in the nQ-bits space
$\mathbb{E}\otimes...\otimes\mathbb{E}$ and defined as
\begin{equation}|\Psi(x_1,...,
x_{n})\rangle=|\psi_{x_1}\rangle\otimes...\otimes|\psi_{x_n}\rangle\end{equation}
where $x_j=0$ or 1 (the number of  such states is obviously
$2^n$). Finally, by using a clever unitary transformation $U$
(details are given in ref.~\cite{PBR}) they found a nice way to
define an orthogonal measurement basis $|\Phi_j\rangle$ (with
$j=1,...,2^n$) in $\mathbb{E}\otimes...\otimes\mathbb{E}$ obeying
to the rule: \\
For every states $|\Psi(x_1,..., x_{n})\rangle $
there exists at least one value of $j$ (this value is different
from one state $|\Psi(x_1,..., x_{n})\rangle $ to one other
$|\Psi(x'_1,..., x'_{n})\rangle $ ) such that
\begin{equation}|\langle\Phi_j|\Psi(x_1,...,
x_{n})\rangle|^2=0.\label{ee}\end{equation}
The basis
$|\Phi_j\rangle$ is actually defined by the  complete set
$U|x'_1,...,x'_{n}\rangle$ and PBR found that for a good choice of
$U$ Eq.~\ref{ee} is satisfied for $x'_1=x_1$,...,$x'_n=x_n$, i.e.,
\begin{equation}
P(\Phi_j|\Psi(x_1,..., x_{n}))=|\langle x_1,...,
x_{n}|U^{\dagger}|\Psi(x_1,...,
x_{n})\rangle|^2=0.\label{eee}\end{equation} We can interpret this
result in the context of $\lambda$-probabilities and write
\begin{widetext}\begin{eqnarray}
P(\Phi_j|\Psi(x_1,..., x_{n}))=\int...\int
\xi(\Phi_j|\lambda_1,...,\lambda_n)\rho_{x_1}(\lambda_1)\cdot
...\cdot\rho_{x_n}(\lambda_n)d\lambda_1....d\lambda_n\nonumber=0\\
\end{eqnarray}\end{widetext} where $\rho_{0}(\lambda)$ and $\rho_{1}(\lambda')$ are  the density
of probability associated with states $|\psi_0\rangle$ and
$|\psi_1\rangle$ respectively. Since these states are independent
we introduced $n$ $\lambda$ variables. It is thus trivial to
repeat the same reasoning as previously: for
$\lambda_1,...,\lambda_n$ belonging to the hypothetical
intersecting support of $\rho_0$ and $\rho_1$  we get
$\xi(\Phi_j|\lambda_1,...,\lambda_n)=0$ for $i=1$ to  $2^n$. Due
to the conservation rule
$\sum_{i}\xi(\Phi_i|\lambda_1,...,\lambda_n)=1$ we obtain the
required PBR contradiction.\\  We finally
deduce the general result:\\
\emph{-PBR Theorem:}\\ \\
 \emph{For any pair of quantum states $\Psi_A$ and $\Psi_B$ in $\mathbb{E}$ the distributions $\rho(\lambda,\Psi_A)$ and $\rho(\lambda,\Psi_B)$ have no common intersecting support. That is we have  $\rho(\lambda,\Psi_A)\cdot\rho(\lambda,\Psi_B)=0$ $\forall \lambda$ in the hidden variable space.}\\
 \\

From this theorem PBR then conclude that the so called
$\Psi$-epistemic ontological models with supplemented   hidden
variable $\lambda$ can not agree with quantum mechanics. Therefore
any hidden variable model must be $\Psi$-ontic  in the sense given
by Harrigan and Spekkens\cite{speckens}.
 \section{Bayes Bell Bohm and PBR}
I think we can  find a simple illustration of what implies   the
PBR theorem. Consider a  50-50 beam splitter and  send a  single
photon state $|\Psi_1\rangle$ through the input  gate 1. The wave
packet split and we will finish with  a probability $P(3|1)=1/2$
to detect the photon in the exit 3 and identically $P(4|1)=1/2$ of
recording  the photon in exit gate 4.  Alternatively, we can
consider a  single photon wave packet coming from gate  2 and at
the end of the photon journey we  will still get
$P(3|1)=P(4|1)=1/2$.  From the point of view of the hidden
variable space we can write
\begin{eqnarray}
   P(4|1 \textrm{ or } 2)=\int \xi(3|\lambda)\rho(\lambda|\Psi_1 \textrm{ or } \Psi_2 )=1/2
\end{eqnarray} with 'or' meaning exclusiveness.
Nothing can be said about the probabilities involved in the
integral. Now, if we consider superposed states such as
$|\pm\rangle=[|\Psi_1\rangle \pm i|\Psi_2\rangle]/\sqrt{2}$ the
photon will finish either in gate 3 or 4 with probabilities
$P(3|+)=P(4|-)=1$ and $P(4|+)=P(3|-)=0$. We here find us in the
orthogonal case of PBR theorem
(i.e. $\langle +|-\rangle=0$). The deduction is thus straightforward and we get  $\rho(\lambda|+ )\rho(\lambda|- )=0$  for all
possible $\lambda$ which means that the two density of probability for superposed states can not have any common intersecting support in the
$\lambda$-space. Nothing to add to this conclusion apparently if we follow PBR.\\
Still, this is I think a not very intuitive result.  Indeed,
spatially $\Psi_1(\mathbf{x})$ and $\Psi_2(\mathbf{x})$ are not
intersecting since they are in two different entrance of the beam
splitter. Therefore in a hidden variable model like the one
proposed by de Broglie-Bohm  (more on this topic is given in the
appendix) where $\lambda$ is the position of the particle
$\mathbf{x}$ in the wave packet we have $\rho(\lambda|\Psi_1
)\rho(\lambda|\Psi_2 )=0$ for all $\lambda$.
This apparently fit quite well with the PBR theorem.\\
However, in this model we  don't have
$\rho(\lambda|+/-)\rho(\lambda|\Psi_2 )=0$ neither we have
$\rho(\lambda|+/-)\rho(\lambda|\Psi_1)=0$ for every $\lambda$!
Indeed, half of the  relevant points of the wave packets $+$ or
$-$ are common to $\Psi_1$ or $\Psi_2$. Actually this is even
worst since we also have
$\rho(\lambda|+)\rho(\lambda|-)=\rho(\lambda|\pm)^2\neq 0$ for
every $\lambda$ in the full $\lambda$-support (sum of the two
disjoint supports associated with  $\Psi_1$ and $\Psi_2$). This is
in complete contradiction with PBR theorem: how could that it
be?\\
I think that PBR, in agreement with Harrigan and Spekkens, would
have qualified the model I am using of $\psi$-ontic in the sense
they are using this word. It means for them that  $\rho(\lambda|+
)\rho(\lambda|- )$  should take a null value which is obviously
not the case. I will give after a detailed account of what is
happening in the de Broglie Bohm model but the main point that I
will try to show now is that we should first (axiomatically) `reject' the
definitions used by Harrigan and Spekkens  as being not general enough (i.e. to make a good classifications of $\lambda$-model) and then
stick to the mathematics to see what PBR missed.\\ In other words
in order to understand the origin of the contradiction we should
work a bit more with the formalism used by PBR to see what is
going on there. For this we go back to the definition of
$\xi(\alpha|\lambda)$ introduced before. Applying naively these
$\xi$ probabilities to our Bohm de Broglie model of the beam
splitter experiment we get $\xi(4|\lambda)=0$, $\xi(3|\lambda)=1$
for every $\lambda$ in the support of $\rho(\lambda|+)$ and
$\xi(3|\lambda)=0$, $\xi(4|\lambda)=1$ for every $\lambda$ in the
support of $\rho(\lambda|-)$.  There is it seems a contradiction
because then this implies $\xi(4|\lambda)=0$ $\xi(3|\lambda)=1$
for every points in the support of $\rho(\lambda|\Psi_1)$ when we
use $+$ and  $\xi(4|\lambda)=0$ $\xi(3|\lambda)=1$ when we use
$-$. Similar contradictions appear on the $\Psi_2$ side. Clearly
there is a problem when one try to use conditional probabilities
such as $\xi(\alpha|\lambda)$ together with the de Broglie Bohm
model.\\
Ok, now lets be a bit more general: we consider the PBR definition
of hidden variable probabilities which for a pure quantum state
$\psi$ generally reads like that:
 \begin{equation}
 |\langle\alpha|\Psi\rangle|^2=P(\alpha|\Psi)=\int \xi(\alpha|\lambda)\rho(\lambda|\Psi)d\lambda
 \end{equation}
where $\alpha$ is the observable eigenvalue associated with the operator $\hat{A}$.  We have also $\sum_{\alpha}\int \xi(\alpha|\lambda)=1$ by definition of a conditional probability. These definitions are very classical like since as we said in the introduction the dynamic or ontic state  should be decoupled from its epistemic counterpart (in agreement with Liouville approach).\\
Now, I remind you the well known Bayes-Laplace probability rule
for two events $\alpha$ and $\beta$:
\begin{eqnarray}
P(\alpha|\beta)P(\beta)=P(\beta|\alpha)P(\alpha)=P(\alpha,\beta).
\end{eqnarray}
Of course for three events $\alpha$, $\beta$ and $u$ we deduce
\begin{eqnarray}
P(\alpha,\beta,u)=P(\alpha,\beta|u)P(u)\nonumber\\
=P(\alpha|\beta,u)P(\beta,u)=P(\alpha|\beta,u)P(\beta|u)P(u)\nonumber\\
\end{eqnarray}
which clearly implies
\begin{eqnarray}
P(\alpha,\beta|u)=P(\alpha|\beta,u)P(\beta|u)
\end{eqnarray}
Now that I reminded you these obvious points I would say that the
most general Bell's hidden variable probability space should obeys
the following rule:
\begin{eqnarray}
P\left(\alpha=\pm 1,\mathbf{a}|\Psi_0 \right)=\int dP\left(\alpha=\pm
1\mathbf{a},\lambda|\Psi_0\right)\nonumber\\=\int P\left(\alpha=\pm
1|\mathbf{a},\lambda,\Psi_0\right)\rho\left(\lambda|\Psi_0\right)d\lambda.\label{e1}
\end{eqnarray}
We eventually used the Heisenberg picture in order to explicitly show the dependency in the initial  quantum state $\Psi_0$. If you don't like conditional probabilities you can alternatively use joint probabilities
\begin{eqnarray}
P\left(\alpha=\pm 1,\mathbf{a},\Psi_0 \right)=\int dP\left(\alpha=\pm
1\mathbf{a},\lambda,\Psi_0\right)\nonumber\\=\int P\left(\alpha=\pm
1|\mathbf{a},\lambda,\Psi_0\right)\rho\left(\lambda,\Psi_0\right)d\lambda.\label{e2}
\end{eqnarray}
In both case $P\left(\alpha=\pm1|\mathbf{a},\lambda,\Psi_0\right)$ plays the role of the $\xi\left(\alpha=\pm1|\mathbf{a},\lambda\right)$ used by PBR. However, now we see the problem: the most general dynamics allowed by the rules of logic should depends on the $\Psi_0$ state considered! \\
Now lets go back to the beam splitter example discussed above. As
shown on Figure 1 here the particle trajectories  in the
$\lambda$-space must be fundamentally different depending on the
choice made for the initial state. This is because
$P\left(\alpha=\pm1|\mathbf{a},\lambda,\Psi_0\right)$ explicitly
depends on $\Psi_0$. The dynamic appears thus clearly different
from the one considered in classical mechanics. Indeed, in a model
like the one proposed by de broglie and Bohm $\Psi$ actually
defines a guiding wave for the particle and is thus an active
partner in the evolution of the $\lambda$-trajectories. Therefore,
we should not be surprised that the trajectories are strongly
influenced \begin{figure}[h]
\begin{center}
\begin{tabular}{c}
\includegraphics[width=8.5cm]{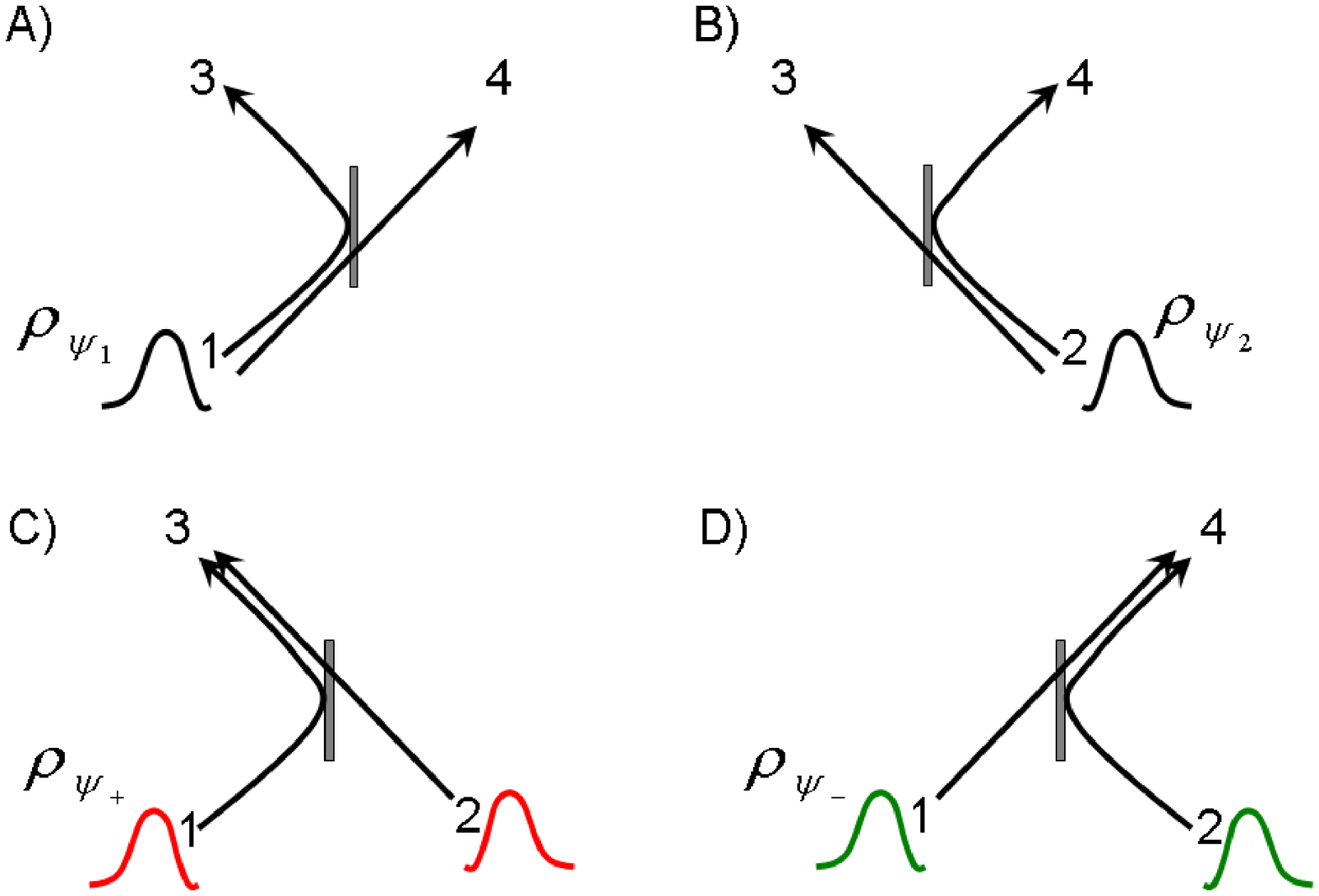}
\end{tabular}
\end{center}
\caption{}
\end{figure}in spatial regions where wave packets interfere or cross. The
beam splitter example is actually reminiscent of the  famous two
slit interference experiment which was treated in details by Bohm
and his followers. The trajectories look sometime `surrealistic' but
this is the price to pay to agree with both a wave and a particle
in a $\lambda$-world. \\
Of course, if we throw away the
$\xi\left(\alpha=\pm1|\mathbf{a},\lambda\right)$ and use instead
$P\left(\alpha=\pm1|\mathbf{a},\lambda,\Psi_0\right)$ the whole
reasoning of PBR collapses since we are not allowed to compare the
states as we did in section 2.\\
Consider for example the orthogonal case. We now have instead of
Eq.~4:
\begin{eqnarray}
|\langle +|\Psi_1\rangle|^2=P(+|\Psi_1)=\int P(+|\lambda,\Psi_1)\rho_1(\lambda)d\lambda=0\nonumber\\
|\langle -|\Psi_2\rangle|^2=P(-|\Psi_2)=\int
P(-|\lambda,\Psi_2)\rho_2(\lambda)d\lambda=0.
\end{eqnarray}
We deduce  of course that $P(+|\lambda,\Psi_1)=0$ if
$\rho_1(\lambda)\neq 0$ and $P(-|\lambda,\Psi_2)=0$ if
$\rho_2(\lambda)\neq 0$. Now If
$\rho_1(\lambda)\cdot\rho_2(\lambda)\neq 0$ for some values of
$\lambda$  (which means once again that $\rho_1$ and $\rho_2$ have
intersecting support) then
\begin{equation}
P(+|\lambda,\Psi_1)=P(-|\lambda,\Psi_2)=0\label{truc}\end{equation}
for the $\lambda$s in the intersection of the two  supports. What
is fundamental here is that contrarily to what occurred for the
models considered by PBR here Eq.~\ref{truc} doesn't imply any
contradiction.  Therefore the PBR theorem cannot be proven any
more! All cases with either orthogonal or non orthogonal states
can always be  analyzed and criticized with the same method:  If
we substitute $\xi(\alpha|\lambda)$ by
$P(\alpha|\lambda,\Psi)$ the PBR theorem can not be proven.\\
The theorem proposed by PBR is thus simply not general enough. It
fits well with the XIX$^{th}$ like hidden variable models but it
is not in agreement with neo-classical model such as the one
proposed by de Broglie and Bohm: QED \emph{reducio ad absurdum}.
In other words: the class of model PBR consider contradict wave
particle duality (see our example with the beam splitter). I think
that peoples who apply naively XIX$^{th}$ century-like epistemic
reasoning to quantum mechanics should seriously worry about PBR
theorem (the others like Bohmian's can sleep peacefully).
Finally, we point out that since Bohm's model is deterministic one
must have
\begin{eqnarray}
P\left(\alpha=\pm1|\mathbf{a},\lambda,\Psi_0\right)=\delta_{\alpha,A(\lambda,\mathbf{a},\Psi_0)}= 0 \textrm{ or } 1
\end{eqnarray}
(where $\delta$ is the Kronecker symbol) since for one given
$\lambda$ only one trajectory is allowed. Equivalently, the actual
value $A(\lambda,\mathbf{a},\Psi_0)=\sum_{\alpha}\alpha
P\left(\alpha=\pm1|\mathbf{a},\lambda,\Psi_0\right)$ can only
takes one of the allowed eigenvalues $\alpha$ associated with the
hermitian operator $A$. We show in the appendix that this is
indeed the case for the particular half-spin model described by
Bohm theory. However the result is actually very general.
\section{Conclusion} Lets be positive: even if PBR theorem is generally wrong it is actually very
interesting: it ruins the old fashion hidden variable approach in
a nice way and show that there are some fundamental differences
between classical $XIX^{th}$ century physics and the neo-classical
mechanics proposed by Bohm and others. Both are based on realism.
Both are admitting an ontic and epistemic parts. But now the wave
function is part of the dynamic all the way along since it gives a
contribution to the ontic state which subsequently affects the
dynamic of the $\lambda$-particle.  The initial Liouville approach
separating the epistemic and the ontic part (i.e. $\rho$ and
$q(t),p(t)$) appears to be wrong if we forget the wave function
(i.e. a same $\lambda$ with different $\Psi_0$ will lead to
different trajectories and density of probability). I think that
PBR managed to do what was the original dream of von Neumann
however both approaches are restricted to a very narrow class of
hidden variable models (which are not orthogonal to each other by
the way).
\appendix
\section{Bohm's deterministic  model for a spin half particle (for those who are not already `Bohred')}
We consider the  simple Q-Bit space for a single spin-$1/2$ \cite{Holland}.
In this model a neutral single particle with spin $1/2$ and mass $M$, is
represented by a wave packet having two components
\begin{eqnarray}\Psi\left(\mathbf{x},t\right)=\left(\begin{array}{l}\psi_{\uparrow}\left(\mathbf{x},t\right)\\
\psi_{\downarrow}\left(\mathbf{x},t\right)\end{array}\right).\end{eqnarray}
In presence of a magnetic field
inside a Stern and Gerlach apparatus the two contributions of  the wave packet are oriented in one or the other of the exits
\cite{Scully,Holland,Holland2}, separating the trajectories
associated with the two states $\uparrow$ and $\downarrow$.
Naturally, any modifications of the magnetic
field orientation change the analyzed basis $\uparrow,\downarrow$.
Consequently in presence of the Stern and Gerlach apparatus
analyzing the spin components along $\mathbf{a}$ and $-\mathbf{a}$
the density of probability
$\rho\left(\mathbf{x},t\right)=|\psi_{\mathbf{a}}\left(\mathbf{x},t\right)|^{2}+|\psi_{-\mathbf{a}}\left(\mathbf{x},t\right)|^{2}$
depends explicitly on the orientation of the magnetic field and must
be written $\rho\left(\mathbf{x},t,\mathbf{a}\right)$.  The evolution  of the wave function in the Stern and Gerlach apparatus is thus given by the pair of equations:
\begin{eqnarray}
i\hbar\partial_{t}\psi_{\mathbf{a}}\left(\mathbf{x},t\right)=-\frac{\hbar^{2}\nabla^{2}}{2M}\psi_{\mathbf{a}}\left(\mathbf{x},t\right)
+\mu (\mathbf{B}(\mathbf{x},t)\cdot\mathbf{a})\psi_{\mathbf{a}}\left(\mathbf{x},t\right)\nonumber\\
i\hbar\partial_{t}\psi_{-\mathbf{a}}\left(\mathbf{x},t\right)=
-\frac{\hbar^{2}\nabla^{2}}{2M}\psi_{-\mathbf{a}}\left(\mathbf{x},t\right)
-\mu(\mathbf{B}(\mathbf{x},t)\cdot\mathbf{a})\psi_{-\mathbf{a}}\left(\mathbf{x},t\right)\label{e3}\nonumber\\
\end{eqnarray}
($\mu$ is the magnetic dipole moment).\\
Now, Bohm says that the ontic state can be described dynamically as a point like object moving with
the velocity
$\mathbf{v}\left(\mathbf{x},t\right)=[\mathbf{J}/\rho]\left(\mathbf{x},t\right)$.
Here
\begin{eqnarray}
\mathbf{J}\left(\mathbf{x},t\right)=\hbar[|\psi_{+\mathbf{a}}\left(\mathbf{x},t\right)|^{2}
\boldsymbol{\nabla}\phi_{+\mathbf{a}}\left(\mathbf{x},t\right)\nonumber\\+|\psi_{-\mathbf{a}}\left(\mathbf{x},t\right)|^{2}
\boldsymbol{\nabla}\phi_{-\mathbf{a}}\left(\mathbf{x},t\right)]/M\nonumber\\
\end{eqnarray}
and
\begin{eqnarray}
\rho\left(\mathbf{x},t\right)=|\psi_{+\mathbf{a}}\left(\mathbf{x},t\right)|^{2}+|\psi_{-\mathbf{a}}\left(\mathbf{x},t\right)|^{2}
\end{eqnarray}
define the probability current and probability density respectively,
and $\phi_{+\mathbf{a}},\phi_{-\mathbf{a}}$ are the phases of
$\psi_{+\mathbf{a}},\psi_{-\mathbf{a}}$.\\
To understand some specificities of this theory  I remind you that from Eqs~\ref{e3} one deduce easily  using the polar form of the wave function
\begin{eqnarray}
-\partial_t |\psi_{\pm\mathbf{a}}\left(\mathbf{x},t\right)|^{2}=-\boldsymbol{\nabla}[|\psi_{\pm\mathbf{a}}\left(\mathbf{x},t\right)|^{2}
\frac{\hbar}{M}\boldsymbol{\nabla}\phi_{\pm\mathbf{a}}\left(\mathbf{x},t\right)]\nonumber\\
\end{eqnarray}
which is the local form of the conservation of probability rule. We also obtain a pair of de Broglie-Bohm version of Hamilton-Jacobi classical equations:
\begin{widetext}
\begin{eqnarray}
-\partial_t \hbar\phi_{\pm\mathbf{a}}\left(\mathbf{x},t\right)=\frac{(\boldsymbol{\nabla}\hbar\phi_{\pm\mathbf{a}}\left(\mathbf{x},t\right))^2}{2M}\pm\mu (\mathbf{B}(\mathbf{x},t)\cdot\mathbf{a})-\frac{\boldsymbol{\nabla}^2|\psi_{\pm\mathbf{a}}\left(\mathbf{x},t\right)|}{2M|\psi_{\pm\mathbf{a}}\left(\mathbf{x},t\right)|}
\end{eqnarray}\end{widetext}
the quantum potential $\frac{\boldsymbol{\nabla}^2|\psi_{\pm\mathbf{a}}|}{2M|\psi_{\pm\mathbf{a}}|}$ is a specific feature of this theory which allows us to describe the quantum statistical properties of the half spin using a classical-like stochastic dynamic. Importantly this quantum potential depends on the  absolute  value of the wave function (up to an arbitrary constant) therefore the dynamical evolution will also depends on the wave function.   This feature is completely different from what occurs in classical physics where the dynamic and the probability are respectively  associated with a pure ontic an epistemic feature.  In classical physics one is free to change the initial density of state without modifying the dynamic. However   here the two features are unseparable since the wave function is part of the ontic and epistemic state at the same time. This feature has a strong consequence on the dynamical evolution which can also be seen more directly  from  the equation of motion
\begin{widetext}
\begin{eqnarray}
\frac{d}{dt}\mathbf{x}(t)=\frac{\hbar[|\psi_{+\mathbf{a}}\left(\mathbf{x},t\right)|^{2}
\boldsymbol{\nabla}\phi_{+\mathbf{a}}\left(\mathbf{x},t\right)+|\psi_{-\mathbf{a}}\left(\mathbf{x},t\right)|^{2}
\boldsymbol{\nabla}\phi_{-\mathbf{a}}\left(\mathbf{x},t\right)]}{M(|\psi_{+\mathbf{a}}\left(\mathbf{x},t\right)|^{2}+|\psi_{-\mathbf{a}}\left(\mathbf{x},t\right)|^{2})}
=\frac{\hbar}{2M}\frac{\textrm{Im}[\Psi^{\dagger}\boldsymbol{\nabla}\Psi]}{\Psi^{\dagger}\Psi}\left(\mathbf{x},t\right).\label{e4}
\end{eqnarray}\end{widetext}
In order to integrate even formally this equation we first have to integrate the Schrodinger equation and we will obtain solution of the form $\psi_{\pm\mathbf{a}}\left(\mathbf{x},t\right)=\int d^{3}\mathbf{x}'K_{\pm\mathbf{a}}\left(\mathbf{x},t,\mathbf{x}',t_0\right)\psi_{\pm\mathbf{a}}\left(\mathbf{x}',t_0\right)$ where  the kernel  $K_{\pm\mathbf{a}}\left(\mathbf{x},t,\mathbf{x}',t_0\right)$ can be evaluated from the Green function. Inserting these solutions in Eq.~\ref{e4} leads to a new  differential equation for $\mathbf{x}(t)$ which reads formally as: \begin{eqnarray}
\frac{d}{dt}\mathbf{x}(t)=\mathbf{G}_{\mathbf{a}}(\mathbf{x}(t),t,\{\Psi^{\dagger}(\mathbf{x}',t_0),\Psi(\mathbf{x}',t_0)\}_{\mathbf{x}'}\forall \mathbf{x}').
\end{eqnarray}
This is a first order equation which not only depends on $\mathbf{x}(t)$ at the  same given time $t$ (i.e. when the derivative is evaluated) but also require the knowledge of the wave function and its complex conjugate evaluated for every position  $\mathbf{x}'$ of the evolution space at the initial time $t_0$. This set of initial values $\{\Psi^{\dagger}(\mathbf{x}',t_0),\Psi(\mathbf{x}',t_0)\}_{\mathbf{x}'}$ plays therefore the role of additional constants of motion. Therefore the complete trajectory  wil be given by a functional having the general form
 \begin{eqnarray}
\mathbf{x}(t)=\mathbf{F}_{\mathbf{a}}(t;\mathbf{x}_0(t),t_0,\{\Psi^{\dagger}(\mathbf{x}',t_0),\Psi(\mathbf{x}',t_0)\}_{\mathbf{x}'}\forall \mathbf{x}').\label{e5}
\end{eqnarray}
Now, in Bohm's model we can define an instantaneous spin vector
\begin{eqnarray}
\mathbf{S}\left(\mathbf{x},t,\mathbf{a}\right)=
\frac{\Psi^{\dagger}\boldsymbol{\sigma}\Psi}{\rho\left(\mathbf{x},t,\mathbf{a}\right)}.
\end{eqnarray} The projection
$\Sigma\left(\mathbf{x},t,\mathbf{a}\right)=\mathbf{S}\left(\mathbf{x},t,\mathbf{a}\right)\cdot\mathbf{a}$
spans a continuum of values during the interaction with the magnetic
field but at end of the measure (i.~e.~at $t=\infty$) we have
$\Sigma=\pm 1$ corresponding to the spin observable $A=\pm1$. We can
naturally define the mean value of the spin projection $\Sigma$ by
\begin{eqnarray}
E_{\Psi}\left(\sigma\right)=\langle\Psi|\boldsymbol{\sigma}\cdot\mathbf{a}|\Psi\rangle=\int
\Sigma\left(\mathbf{x},t,\mathbf{a}\right)\rho\left(\mathbf{x},t,\mathbf{a}\right)d^{3}\mathbf{x}.\nonumber\\
\end{eqnarray}
We can always define univocally the
actual position $\mathbf{x}\left(t\right)$ measured for example at
$t=+\infty$ by a function of the initial coordinate
$\mathbf{x}_{0}=\lambda$ of the particle at a time $t_{0}\rightarrow
-\infty$, i.~e.~a long time before that the particle enters in the
Stern and Gerlach apparatus. Due to the conservation of probability requirement
the number of states defined by
$\rho\left(\mathbf{x}_{0},t_{0}\right)\delta^{3}\mathbf{x}_{0}$ in
the elementary volume $\delta^{3}\mathbf{x}_{0}$ is naturally
identical to
$\rho\left(\mathbf{x}\left(t\right),t,\mathbf{a}\right)\delta^{3}\mathbf{x}\left(t\right)$
i.~e.~:
\begin{eqnarray}
\rho\left(\mathbf{x}\left(t\right),t,\mathbf{a}\right)\delta^{3}\mathbf{x}\left(t\right)
=\rho\left(\mathbf{x}_{0}(t_0),t_{0}\right)\delta^{3}\mathbf{x}_{0}\left(t_0\right).
\end{eqnarray}
This result is of course well known in fluid dynamics where it is associated
to the names of Euler and Lagrange (the so called Euler-Lagrange
coordinates). This law  can also be written
\begin{widetext}\begin{eqnarray}
\int_{\delta V}\rho\left(\mathbf{x}',t,\mathbf{a}\right)d^{3}\mathbf{x}'=\int [ \int_{\delta V}\delta^{3}(\mathbf{x}'-\mathbf{x}(t))d^{3}\mathbf{x}']\rho\left(\mathbf{x}(t),t,\mathbf{a}\right)d^{3}\mathbf{x}(t)\nonumber\\
=\int [\int_{\delta V}\delta^{3}(\mathbf{x}'-\mathbf{F}_{\mathbf{a}}(t;\mathbf{x}_0(t),t_0,\{\Psi^{\dagger}(\mathbf{x}',t_0),\Psi(\mathbf{x}',t_0)\}_{\mathbf{x}'}\forall \mathbf{x}'))d^{3}\mathbf{x}']\rho\left(\mathbf{x}_0(t_0),t_0\right)d^{3}\mathbf{x}_0(t_0)
\end{eqnarray}\end{widetext}
The second line in this equation is deduced from Eq.~\ref{e5}. This expression is therefore a generalization  for the continuous observable $\mathbf{x}$ of Eq.~18 which is valid only for dichotomic observable.
Similarly
$\Sigma\left(\mathbf{x}\left(t\right),t,\mathbf{a}\right)$ can be
expressed as a function of the initial coordinates of the particle
and can be written
$A\left(\mathbf{x}_{0}(t_0),t_{0},\{\Psi^{\dagger}(\mathbf{x}',t_0),\Psi(\mathbf{x}',t_0)\}_{\mathbf{x}'}\forall \mathbf{x}',t,\mathbf{a},\right)$.  If we consider now
the expectation value
$\langle\Psi|\boldsymbol{\sigma}\cdot\mathbf{a}|\Psi\rangle$, we can
write\begin{widetext}
\begin{eqnarray}
E_{\Psi}\left(\sigma\right)=\langle\Psi|\boldsymbol{\sigma}\cdot\mathbf{a}|\Psi\rangle=\int\Sigma\left(\mathbf{x},t,\mathbf{a}\right)\rho\left(\mathbf{x},t,\mathbf{a}\right)d^{3}\mathbf{x}\nonumber\\
=\int A\left(\mathbf{x}_{0}(t_0),t_{0},\{\Psi^{\dagger}(\mathbf{x}',t_0),\Psi(\mathbf{x}',t_0)\}_{\mathbf{x}'}\forall \mathbf{x}',t,\mathbf{a}\right)\rho\left(\mathbf{x}_{0},t_{0}\right)d^{3}\mathbf{x}_{0}(t_0).
\end{eqnarray}\end{widetext}
If we choose $t=+\infty$ then $A=\pm 1$ and we have the complete
definition of Bell (with now $\rho(\lambda)$ independent of
$\mathbf{a}$ as desired).\\
One can also define\begin{widetext}
\begin{eqnarray}
\mathcal{P}(\alpha=\pm 1,\mathbf{a})=\frac{1\pm\langle\Psi|\boldsymbol{\sigma}\cdot\mathbf{a}|\Psi\rangle}{2}=\int\frac{(1\mp\Sigma\left(\mathbf{x},t,\mathbf{a}\right))}{2}\rho\left(\mathbf{x},t,\mathbf{a}\right)d^{3}\mathbf{x}\nonumber\\
=\int\frac{(1\pm A\left(\mathbf{x}_{0}(t_0),t_{0},\{\Psi^{\dagger}(\mathbf{x}',t_0),\Psi(\mathbf{x}',t_0)\}_{\mathbf{x}'}\forall \mathbf{x}',t,\mathbf{a}\right))}{2}\rho\left(\mathbf{x}_{0},t_{0}\right)d^{3}\mathbf{x}_{0}(t_0).
\end{eqnarray}\end{widetext}
This quantity approaches  asymptotically the definition of the projector operator on the $\pm\mathbf{a}$ direction and therefore gives us the probability for the dichotomic spin projection observable. The quantity
\begin{eqnarray}
\frac{(1\pm A)}{2}=\delta_{\alpha=\pm1,A}=0\textrm{ or } 1
\end{eqnarray}
is indeed the conditional probability
$P_{1}\left(\alpha=\pm1|\mathbf{a},\lambda,\Psi\right)$ discussed
in the manuscript (see Eq.~18).

\end{document}